# Comparing the Predictions of two Mixed Neutralino Dark Matter Models with the Recent CDMS II Candidate Events

D. P. Roy

*Homi Bhabha Centre for Science Education, Tata Institute of Fundamental Research,
V. N. Purav marg, Mumbai-400088, India*

We consider two optimally mixed neutralino dark matter models, based on nonuniversal gaugino masses, which were recently proposed by us to achieve WMAP compatible relic density over a large part of the MSSM parameter space. We compare the resulting predictions for the spin-independent DM scattering cross-section with the recent CDMS II data, assuming the possibility of the two reported candidate events being signal events. For one model the predicted cross-section agrees with the putative signal over a small part of the parameter space, while for the other the agreement holds over the entire WMAP compatible parameter space of the model.

**Introduction:** Recently the CDMS II experiment has published the results of their final data on dark matter (DM) scattering on Germanium nucleus, showing two events in the signal region with recoil energies of 12.3 keV and 15.5 keV [1]. Their estimated surface electron background over the signal region is $0.8\pm0.1(stat)\pm0.2(syst)$. Adding the small neutron background to this gives an overall probability to observe $\geq 2$ background events in the signal region as 23%, which is not very small. Therefore they have concluded that this result cannot be interpreted as significant evidence for DM signal, but they cannot reject either event as signal. In view of the former they have only given a 90% CL upper limit on the signal cross-section assuming the two candidate events to be signal events. In view of the latter observation, however, over a dozen of phenomenological papers have already appeared proposing as many new DM models to explain a possible DM signal, represented by the two candidate events. It should be added here that, because of the two candidate signal events, their combined 90% CL upper limit from present plus earlier data [2] is not much stronger than that obtained from the earlier data alone. In any case both these upper limits are well above the predicted signal rates of most DM models. Thus their upper limit by itself does not call for any new DM model. However, to the extent that the two candidate events (or at least one of them) have a good chance of being a DM signal, it is worth identifying potential DM models that can naturally explain such a signal.

For any meaningful comparison of a DM model prediction with a presumed signal, represented by the two candidate events of the CDMS II experiment, one needs to know the central value and error corridor of the signal cross-section, corresponding to these events. For this purpose we shall assume the standard Poisson distribution, for which the 90% CL upper and lower limits for 2 signal events over a low background are 5.3 and 0.53 events respectively [3]. The CDMS II paper [1] has used instead the Optimal Interval Method [4] to compute their 90% CL upper limit for the 2 signal events. However, we have checked from ref [4] that the 90% CL upper limits evaluated with the Poisson and the Optimal Interval Methods agree to within 10-15%. We prefer to use the Poisson method because it is more standard and one can simply read off the 90% CL upper and lower limits from the PDG [3]. Thus we shall estimate the central value and the 90% CL lower limits of the presumed signal cross-section by simply scaling down the 90%CL upper limit curve of the CDMS II paper [1] by factors of 2/5.3 and 1/10 respectively. Admittedly these estimates hang on the optimistic assumption of the 2 candidate events being signal events. But no meaningful comparison with model predictions for DM signal is possible without such an assumption. Alternatively one may assume at least 1 of the 2 candidate events to be a signal event, which would scale down the 90% CL lower limit by a factor of 5 [3]. One can easily see the effect of this change on our model predictions. Of course, a definitive answer to whether there is a DM signal at the level implied by these candidate events will have to await the result from the superCDMS experiment [5], which is expected in a few years time. Till then one can only do a provisional analysis by assuming the signal to be at the level suggested by these events.

In this note we compare the presumed signal cross-section, corresponding to these candidate events, with the predictions of two mixed neutralino DM models, recently suggested by us [6] to achieve cosmologically compatible relic density over a large range of the MSSM parameters. They are based on the assumption that supersymmetry is broken by an admixture of two superfields, belonging to a singlet and a nonsinglet representation of the GUT group, leading to nonuniversal gaugino masses at the GUT scale [7]. For the simplest case of SU(5) GUT, we could construct two models corresponding to (1+75) and (1+200) representations of the SUSY breaking superfields. In each case, the relative size of the singlet and nonsinglet contributions was adjusted to give a large admixture of the gaugino and higgsino components in the neutralino DM, so as to achieve a WMAP satisfying relic density over large range of the model parameters. Each model automatically predicts a large spin-independent (SI) cross-section for DM scattering on nucleon, leading to a promising DM signal for direct detection experiments like CDMS. There is no overlap of these models with the nonuniversal gaugino mass models constructed recently to explain these candidate events [8].

The following section summarizes the essential steps in the construction of the (1+ 75) and (1 +200) models. In the next section we compare the predictions of the two models with the putative DM signal, indicated by the two candidate events. We find agreement of the signal with the (1 +75) model prediction only over a small part of the parameter space. On the other hand, the agreement with the (1 +200) model prediction holds over the entire WMAP satisfying parameter space of the model. For both cases we discuss the resulting SUSY spectra and their implications for the SUSY signals at LHC.

**The (1 +75) and (1 + 200) Models for the Mixed Neutralino DM :** The gauge kinetic function responsible for the GUT scale gaugino masses arises from the vacuum expectation value of the F-term of a chiral superfield Φ, which is responsible for SUSY breaking, i.e.

$$\frac{\langle F_\Phi \rangle_{ij}}{M_{Plank}} \lambda_i \lambda_j, \tag{1}$$

where $\lambda_{1,2,3}$ are the U(1), SU(2) and SU(3) gaugino fields – bino, wino and gluino. Since the gauginos belong to the adjoint representation of SU(5), $\Phi$ and $F_\Phi$ can belong to any of the irreducible representations appearing in their symmetric product [7], i.e.

$$(24 \times 24)_{symm} = 1 + 24 + 75 + 200. \tag{2}$$

Thus the GUT scale gaugino masses in a given representation $n$ are determined in terms of a single SUSY breaking mass parameter by

$$M^G_{1,2,3} = C^n_{1,2,3} m^n_{1/2}, \tag{3}$$

where

$$C^1_{1,2,3} = (1,1,1); C^{24}_{1,2,3} = (-1,-3,2); C^{75}_{1,2,3} = (-5,3,1); C^{200}_{1,2,3} = (10,2,1). \tag{4}$$

The minimal SUGRA model assumes $\Phi$ to be a singlet, leading to universal gaugino masses at the GUT scale. On the other hand any of the nonsinglet representations for $\Phi$ would imply nonuniversal gaugino masses as per eqs. (3) and (4). The phenomenology of such nonuniversal gaugino mass models have been extensively studied in the literature [9]. Since the gaugino masses evolve like the corresponding gauge couplings at the one-loop level of the RGE, the three gaugino masses at the electroweak (EW) scale are proportional to the corresponding gauge couplings, i.e.

$$\begin{aligned} M_1 &= (\alpha_1/\alpha_G) \simeq (25/60) C^n_1 m^n_{1/2}, \\ M_2 &= (\alpha_2/\alpha_G) \simeq (25/30) C^n_2 m^n_{1/2}, \\ M_3 &= (\alpha_3/\alpha_G) \simeq (25/9) C^n_3 m^n_{1/2}. \end{aligned} \tag{5}$$

The higgsino mass parameter µ is obtained from the EW symmetry breaking condition and the one-loop RGE for the Higgs scalar mass, i.e.

$$\mu^2 + M_Z^2/2 \simeq -m^2_{H_u} \simeq -0.1 m_0^2 + 2.1 M_3^{G^2} - 0.22 M_2^{G^2} - 0.19 M_2^G M_3^G, \tag{6}$$

neglecting the contribution from the GUT scale trilinear coupling term $A_0$. The numerical coefficients on the right correspond to a representative value of tan β = 10; but they show only mild variation over the moderate tan β region.

Although we use exact numerical solutions to the two-loop RGE in our analysis, the composition of the lightest neutralino DM $\chi^0_1$ (abbreviated as χ) can be seen from the relative values of the gaugino and higgsino masses, given by eqs. (3-5) and eq. (6) respectively. For the universal gaugino mass model (singlet $\Phi$), one gets $M_1 < \mu$, resulting in a bino DM over most of the parameter space. Since bino has no gauge charge, it can only pair annihilate via sfermion exchange; and the large sfermion mass limit from LEP [3] leads to overabundance of DM relic density. The same is true for the 24-plet representation. On the other hand, for the 75 and 200-plet representations, one gets $M_{1,2} > \mu$, resulting in a higgsino DM over most of the parameter space. And since the higgsino DM can co-annihilate efficiently with its nearly degenerate chargino via W-boson, one gets underabundance of DM relic density [10]. Finally, assuming the SUSY breaking to occur via an admixture of a singlet and a nonsinglet superfield belonging to the (1 +75) or (1 +200) representations [11], i.e.

$$m_{1/2}^1 = (1-\alpha_{75})m_{1/2} \ \& \ m_{1/2}^{75} = \alpha_{75}m_{1/2},$$
$$m_{1/2}^1 = (1-\alpha_{200})m_{1/2} \ \& \ m_{1/2}^{200} = \alpha_{200}m_{1/2},$$
(7)

one can get a large admixture of bino and higgsino components in the DM by adjusting the mixing parameter α. We got optimal admixture of bino and higgsino components in the (1 +75) model and bino, wino and higgsino components in the (1 +200) model [6] with

$$\alpha_{75} = 0.475 \ \& \ \alpha_{200} = 0.12,$$
(8)

leading to WMAP [12] satisfying relic densities over large parts of the MSSM parameter space. These are two simple realizations of the so called well-tempered neutralino scenario [13]. Once the mixing parameter is fixed, each of these models is as predictive as the minimal SUGRA model.

**Comparing the Predictions of the (1 + 75) and (1 + 200) Models with the Signal Level Suggested by the Candidate Events:** The direct DM detection signal on Germanium is based on the elastic scattering of DM on Ge nucleus, which is dominated by the spin-independent (SI) interaction due to Higgs exchange [14]. Since the Higgs coupling to the DM pair is proportional to the product of their higgsino and gaugino components, these mixed neutralino DM models predict rather large SI cross-sections and hence promising signals for the direct detection experiments like CDMS. We shall compare the model predictions with the putative signal cross-section corresponding to the CDMS II candidate events, first for the (1 +75) and then the (1 + 200) model. The results are fairly stable over the intermediate tan β region in both cases.

Figure 1 compares the (1 + 75 ) model prediction for the SI elastic scattering cross-section of DM on nucleon with the putative signal, corresponding to the two candidate events of the CDMS II experiment [1]. The signal corridor is described by the central value and the 90% CL upper and lower limits of the cross-section, which were obtained using the Poisson method described above. The signal corridor is seen to select only a small corner of the model parameter space, corresponding to a small DM mass ≈ 100 GeV. The SUSY spectrum corresponding to this region is listed in Table 1. A characteristic feature of this mixed bino-higgsino DM model is the near degeneracy of lighter chargino and neutralino masses. This implies rather soft leptons from the SUSY cascade decay along with a hard missing-$E_T$ carried away by the DM. Thus one expects a robust missing-$E_T$ signature for this model at LHC. The low value of gluino mass implies that one can see this signal even at the first stage of the 10 TeV LHC run with a low luminosity of about 100 pb$^{-1}$ [15].

Table 1. Superparticle masses (in GeV) for a WMAP compatible point in the intersection region of the (1 + 75) model prediction with the CDMS II candidate events of Fig 1, corresponding to $m_{1/2}$ = 144 GeV and $m_0$ = 1255 GeV. All the remaining sfermion and Higgs boson masses are around 1250 GeV.

| $\chi^0_1(\chi)$ | $\chi^0_2$ | $\chi^0_3$ | $\chi^0_4$ | $\chi^+_1$ | $\chi^+_2$ | $\tilde{g}$ | $\tilde{t}_1$ | $\tilde{t}_2$ | $\tilde{b}_1$ | $h^0$ |
|---|---|---|---|---|---|---|---|---|---|---|
| 103 | 120 | 168 | 270 | 121 | 270 | 433 | 760 | 1063 | 1054 | 112 |

Figure 2 compares the analogous prediction of the (1 + 200) model with the signal corridor, corresponding to the two candidate events. Here we see that the entire WMAP compatible parameter space falls within the signal corridor. The SUSY spectra for two representative points over this WMAP compatible parameter space, which lie very close to the central value curve for the two candidate events, are listed in table 2. Let us summarize the main features. i) The large bino-wino-higgsino mixing leads to an approximate degeneracy between all the electroweak charginos and neutralinos. ii) The gluino is about two and half times heavier. iii) There is an inverted hierarchy in the squark sector – the lighter top squark is lighter than the gluino, while other squarks are heavier or at least as heavy as the gluino. iv) The WMAP compatible region corresponds to a relatively heavy DM mass of 400-900 GeV, with the corresponding gluino mass of 1.0-2.3 TeV [6].

Table 2. SUSY breaking mass parameters and superparticle masses (in GeV) for two representative points in the WMAP compatible parameter space of the (1 + 200) model, which lie close to the central value curve for the CDMS II candidate events of Fig 2. All the remaining squark masses are in the range of 2000-2240 GeV.

| $m_{1/2}$ | $m_0$ | $\chi^0_1(\chi)$ | $\chi^0_2$ | $\chi^0_3$ | $\chi^0_4$ | $\chi^+_1$ | $\chi^+_2$ | $\tilde{g}$ | $\tilde{t}_1$ | $\tilde{t}_2$ | $\tilde{b}_1$ | $h^0$ |
|---|---|---|---|---|---|---|---|---|---|---|---|---|
| 725 | 1450 | 633 | 657 | 794 | 822 | 643 | 818 | 1700 | 1460 | 1813 | 1801 | 117 |
| 900 | 1357 | 798 | 818 | 985 | 1009 | 807 | 1005 | 2045 | 1649 | 2013 | 2001 | 118 |

The inverted hierarchy implies a large number of top quarks from the decay of the gluino pair at LHC in addition to the large missing-$E_T$, carried by the DM pair. This leads to a final state with 3-4 top quarks along with a large missing-$E_T$, analogous to the case of the focus point region [16]. Thus, as in the case of the focus point region, one expects a distinctive LHC signature, with multiple isolated leptons and b quarks from top decays along with a large missing-$E_T$. In order to probe the gluino mass range up to 2.3 TeV, however, one would need a luminosity of 100 fb$^{-1}$ at 10 TeV (or 10 fb$^{-1}$ at 14 TeV) [15].

It should be added here that, if the DM candidate events are confirmed by the superCDMS [5] experiment with a 10 times higher statistics, one can then use the more precise signal level to distinguish the parameter regions mapped by the three lines of Figure 2. Given that the two candidate events come from two years data of CDMS II with 5 kg of Ge, the superCDMS with 15 kg of Ge will give 10 times higher statistics after seven years of data taking. Hopefully by then one would have mapped the SUSY spectrum at LHC; and so one can make a quantitative comparison between the SUSY parameters selected by the two experiments.

Finally, some of the abovementioned theoretical papers have claimed that the low nuclear recoil energy of the two candidate events of CDMS II imply a low DM mass of $\leq$ 100 GeV. However, even for a relatively high DM mass the recoil nuclear energy spectrum is expected to decline rapidly because of the form factor effects [14, 17]. Thus, while a large recoil energy implies a high DM mass, the converse is not necessarily true. Nonetheless, it should be noted from Figure 2 that in the low DM mass region of around 100 GeV, the (1 + 200) model predicts a 4-5 times larger cross-section than the putative signal. Interestingly, it also predicts too large an annihilation cross-section in this region, resulting in an underabundance of DM relic density by a similar factor [6]. Therefore, assuming a two-component DM scenario, where the SUSY DM accounts for a fraction f of the relic density in the region of underabundance, one should scale down the model prediction in this region by the fraction f. In this way one can simultaneously reconcile the (1 + 200) model predictions with both the WMAP relic density and the putative direct detection signal in the low DM mass region as well.

I thank Manuel Drees and Utpal Chattopadhyay for discussions, and also the latter for help with the figures. The work was supported in part by the DAE Raja Ramanna fellowship.

## References

[1] Z. Ahmed et al. [CDMS Collaboration], arXiv:0912.3592 [astro-ph.CO], submitted to Science.
[2] Z. Ahmed et al. [CDMS Collaboration], Phys. Rev. Lett. **102**, 011301 (2009).
[3] C. Amsler et al. [Particle Data Group], Phys. Lett. B **667**, 1 (2008).
[4] S. Yellin, Phys. Rev. D **66**, 032005 (2002).
[5] J. Cooley for CDMS II Collaboration, arXiv:0912.1601 [astro-ph.CO].
[6] U. Chattopadhyay, D. Das and D. P. Roy, Phys. Rev. D **79**, 095013 (2009).
[7] J. Ellis et al. Phys. Lett. B **155**, 381 (1985); M. Drees, Phys. Lett. B **158**, 409 (1985).
[8] M. Holmes and B. D. Nelson, arXiv:0912.4507 [hep-ph]; I. Gogoladze et al., arXiv:0912.5411 [hep-ph]; A. Bottino et al., arXiv:0912.4025 [hep-ph].
[9] G. Anderson, H. Baer, C. H. Chen and X. Tata, Phys. Rev. D **61** 095005 (2000); K. Huitu, Y. Kawamura, T. Kobayashi and K. Poulamaki, Phys. Rev. D **61**, 03500 (2000); U. Chattopadhyay and P. Nath, Phys. Rev. D **65**, 075009 (2002).
[10] U. Chattopadhyay and D. P. Roy , Phys. Rev. D **68**, 033010 (2003).
[11] S. F. King, J. P. Roberts and D. P. Roy, JHEP 0710:106 (2007).
[12] E. Komatsu et al. [WMAP Collaboration], Astrophys. J. Suppl. Ser. **180**, 330 (2009).

[13] N. Arkani-Hamed, A. Delgado and G. F. Giudice, Nucl. Phys. B **741**, 108 (2006).
[14] G. Jungman, M. Kamionkowski and K. Griest, Phys. Rept. **267**, 195 (1996);
    J. D. Lewin and P. F. Smith, Astropart. Phys. **6**, 87 (1996).
[15] H. Baer, V. Barger, A. Lessa and X. Tata, JHEP 0909:063 (2009).
[16] U. Chattopadhyay, A. Datta, A. Datta, A. Datta and D. P. Roy, Phys. Lett. B **493**, 127 (2000); P. Mercandate, J. Mizokoshi and X. Tata, Phys. Rev. D **72**, 035009 (2005)**;** H. Baer et al., Phys. Rev. D **75**, 095010 (2007).
[17] M.Drees and C. Shan, astro-ph/0703651.

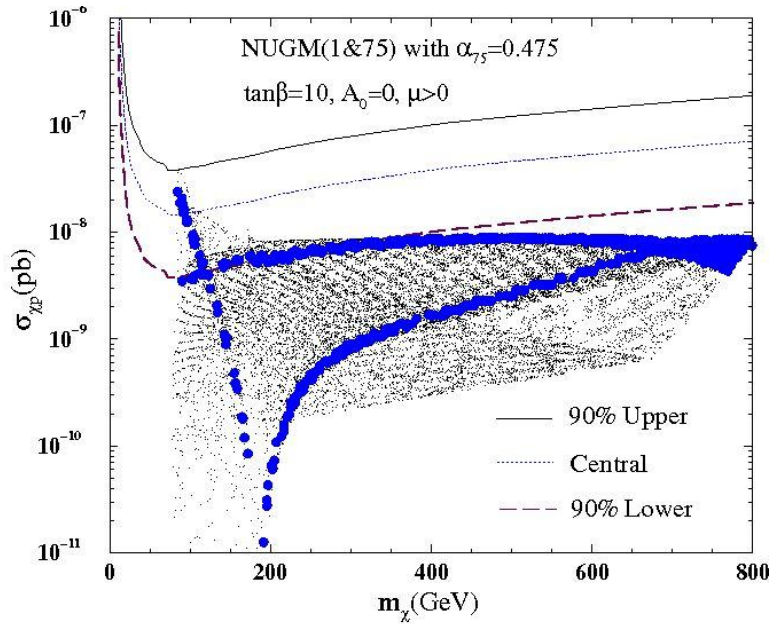

Fig 1. Prediction of the (1+ 75) model compared with the putative signal corridor, corresponding to the two candidate DM scattering events of the CDMS II experiment [1]. The blue (dark) dots correspond to the WMAP compatible DM relic density region.

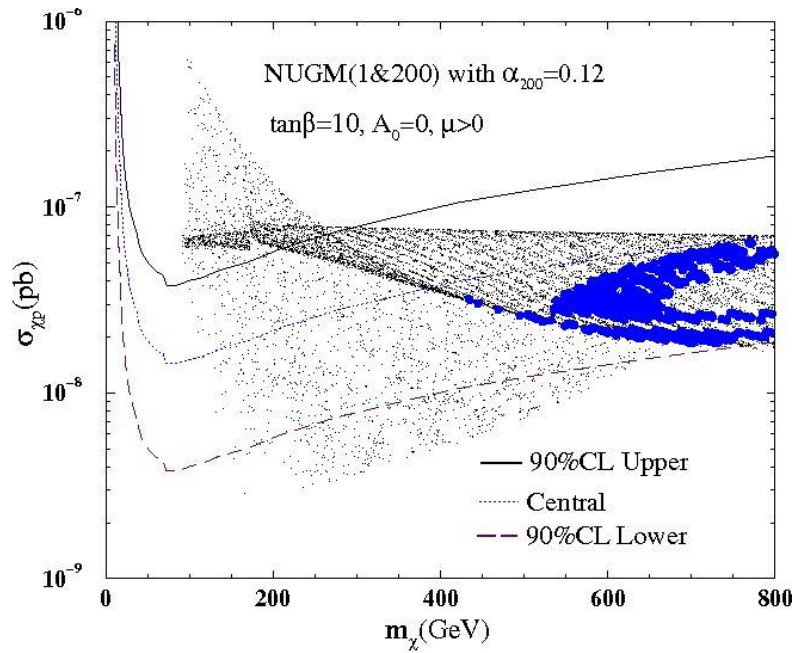

Fig 2. Same as Fig 1 for the (1+ 200) model.